\documentclass[11pt,twoside]{article}

\usepackage{asp2006}
\usepackage{epsf}
\usepackage{lscape}
\usepackage{graphicx}

\begin{document}
\title{Variability Study of EHB Stars in the Globular Cluster NGC~6752}   
\author{Janusz Kaluzny}   
\affil{Nicolaus Copernicus Astronomical Center, Bartycka 18, 00-716 Warsaw,
Poland}    

\author{Ian Thompson}   
\affil{Observatories of the Carnegie Institution of Washington, 813 Santa Barbara Street, 
Pasadena, California 91101}    

\begin{abstract} 
We present the results of a search for variable stars in the 
central part of the globular cluster NGC~6752. The monitored sample included 
160 BHB and 107 EHB stars, respectively. A total of 17 variables were 
detected of which 14 are new identifications. Five variables are 
BHB/EHB stars. We report also on identification of a detached
eclipsing binary being likely a member of the cluster. Moreover, we 
detected an outburst of a dwarf nova located in the cluster core.  
\end{abstract}


\section{Introduction}
NGC~6752 is a nearby post core-collapsed globular cluster. 
In his compilation \citet{harris} lists for it
$(m-M)_{\rm V}=13.02$ and  $E(B-V)=0.04$. The cluster harbors a rich population
of blue horizontal branch (BHB) and extreme horizontal branch (EHB) stars
\citep{buonanno, momany}. The photometric survey conducted with the 
1-m Swope telescope by the CASE group \citep{thompson}  resulted in 
detection of 3 SX~Phe stars and 8 eclipsing binaries in the 
cluster field. So far none of the blue stars in NGC~6752 is a known 
photometric variable. This is a bit unexpected in light of the idea that 
most or even all EHB stars are components of close 
binary systems \citep{mengel, heber02}.
Close binaries are common among field EHB stars \citep{maxted, napiwotzki}.
On the other hand \citet{bidin06} detected no spectroscopic binaries among
51 BHB/EHB stars in NGC~6752. Further evidence for a lack of close binaries 
among EHB stars in globular clusters was presented during this conference
by Moni Bidin.\\ 
In this contribution we present results of a new survey for 
photometric variables in NGC~6752. It covers smaller area than
our earlier work but is deeper and better suited for study of
the innermost region of the cluster. 

\section{Observations and Reductions} 

The central part of the cluster was observed with the 2.5-m DuPont
telescope during 8 consecutive nights in May 1998. The 
2048$^2$ TEK5 CCD camera was used as a detector with a scale of 
0.26 arcsec/pixel and field of view of $8.8\times 8.8$ arcmin$^2$.
The cluster was monitored for a total of 30 hours with $B$ \& $V$ 
filters. The average exposure times were 35s and 60s for $V$ and $B$
filters, respectively. Such rather short exposure times helped
to avoid saturation of too many stars in the dense core region of 
the cluster. The sequences of 2-4 individual images were combined to improve
the signal to noise ratio and to speed up data reductions. 
The results presented here are based on 152 combined images
in $V$ and 143 combined images in $B$. Effective resolution 
of the data ranges from 4 to 8 minutes. \\
Photometric reductions were conducted using Daophot/Allstar
\citep{stetson} and ISIS \citep{alard} packages.
The light curves for about 30 000 stars were extracted and analyzed for 
variability using AoV and AOVTRANS programs \citep{alex, alex1}.

\section{Results}
We detected 17 variable stars of which 14 are new identifications.
Location of 15 of them is marked on the color magnitude-diagram 
of the cluster shown in Fig. 1. 
%
%

\begin{figure}[!ht]
\begin{center}
\includegraphics[scale=0.6]{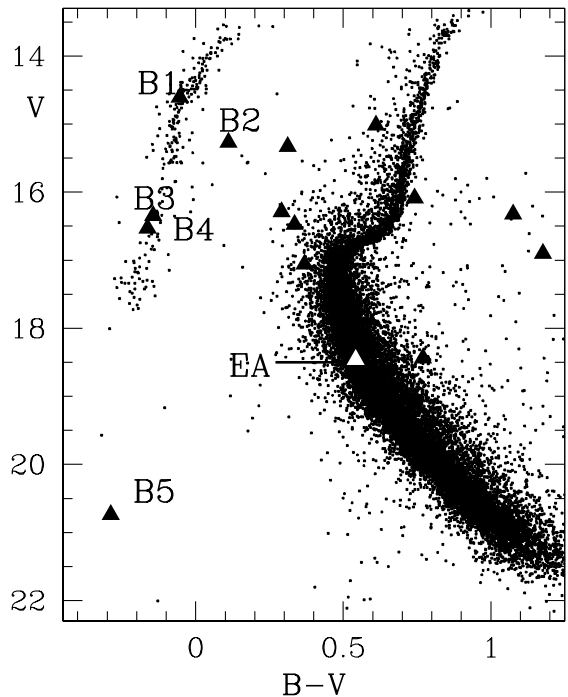}
\end{center}
\caption{Color-magnitude diagram of NGC~6752 with marked variables.}\label{fig1}
\end{figure}

\begin{figure}[!ht]
\begin{center}
\includegraphics[scale=0.6]{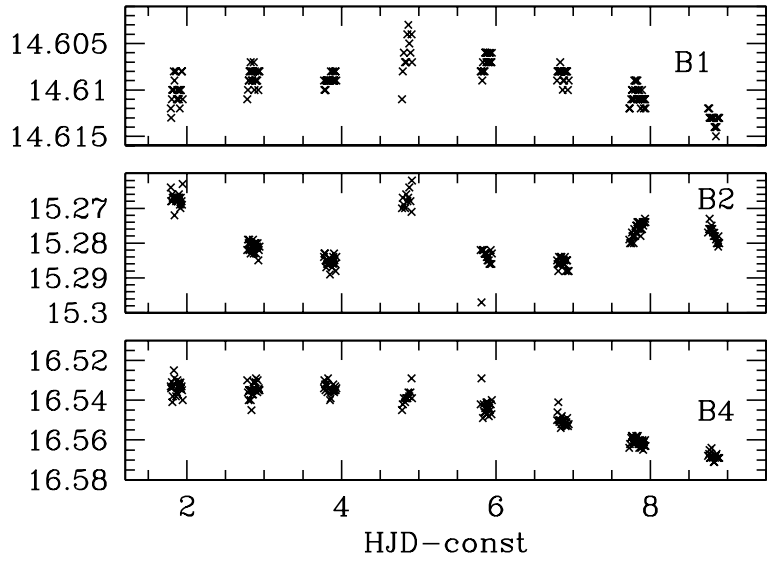}
\end{center}
\caption{V-band light curves of variables B1-2 and B4.}\label{fig2}
\end{figure}

\begin{figure}[!ht]
\begin{center}
\includegraphics[scale=0.65]{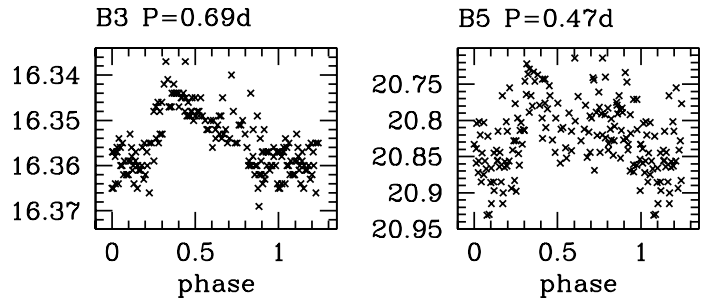}
\end{center}
\caption{V-band light curves of variables B3 and B5.}\label{fig3}
\end{figure}

\begin{figure}[!ht]
\begin{center}
\includegraphics[scale=0.65]{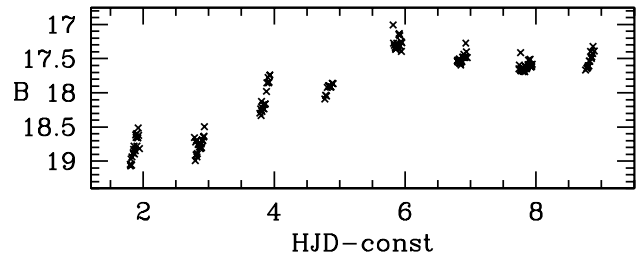}
\end{center}
\caption{B-band light curve of a candidate dwarf nova.}\label{fig4}
\end{figure}

Stars B1 and B3-5 are candidate BHB/EHB
objects. As for variable B2 it is a likely binary containing BHB/EHB star
and a redder companion. Time domain light curves of B1, B2 and B4 
are shown in Fig. 2. Light curve of B2 shows likely periodicity 
with $P\approx 3.3$~d or $P\approx 6.6$~d. It is possible that 
all 3 objects are binaries and that their variability is due
to ellipsoidal and/or reflecting effect. Light curves of B3 and B5 
can be phased with periods $P=0.69$~d and $P=047$~d, respectively.
Their phased light curves are shown in Fig. 3. 
The light curve of B3 is asymmetric which suggests that it is
a pulsating variable. However, the observed period is too long as 
for pulsating sdB star. As for B5 we propose, based on its faintness
and  observed period, that it is related to cataclysmic variables.
We note parenthetically that B5 is also a UV bight object 
with $U-B\approx -1.0$. As for the remaining variables we would like to 
higlight a detection of a detached eclipsing binary which is located on the 
cluster main sequence. On Fig. 1 this variable is labeled as EA.
We detected only one eclipse for it with a depth of about 0.2 mag in the V-band.
The eclipse seems to be total. Detailed analysis of this object can provide
a direct determination of absolute parameters for two main sequence 
stars belonging to the cluster. Finally, we report on a detection of a
possible dwarf nova outburst for an object located right
at the cluster core. Only $B$ band light curve could be extracted
for it. Photometry in the $V$ band was impossible due to a presence of
a nearby bright stars with saturated images. The light curve of 
the candidate dwarf nova is shown in Fig. 4.\\ 
\citet{pooley} reported  detection of several X-ray sources in 
the core of NGC~6752. It remains to be checked if any of these 
sources corresponds to the object which showed an optical outburst.  
More detailed analysis of all detected variables will be presented soon.
In particular we plan an extraction of light curves based on individual images
rather than on stacked frames. 
Moreover, we are aiming at spectroscopic follow up of selected variables.
Spectroscopic data will allow in particular to check on the binary
nature of detected blue variables.

\acknowledgements 
JK was supported by the grant 1~P03D~001~28 from the Ministry
of Science and Higher Education, Poland.
IBT was supported by NSF grant AST-0507325.



\begin{thebibliography}{}
\bibitem[Alard \& Lupton(1998)]{alard}
 Alard, C., \& Lupton, R.~H.  1998, ApJ, 503, 325

\bibitem[Buonanno et al. (1986)]{buonanno}Buonanno, R., Caloi, V., 
Castellani, V., Corsi, C., Fusi Pecci, F., \& Gratton, R. 1986, A\&AS, 66, 79	

\bibitem[Harris(1996)]{harris}
Harris, W.~E. 1996, AJ, 112, 1487

\bibitem[Heber et al. (2002)]{heber02}Heber, U., Moechler, S., Napiwotzki, R., 
Thejll, P., \& Green, E.~M. 2002, A\&A, 383, 938

\bibitem[Maxted et al. (2001)]{maxted}Maxted, P.~F.~L., Heber, U., 
Marsh, T.~R., \& North, R.C. 2001, MNRAS, 326, 1391

\bibitem[Mengel, Norris, \& Gross (1976)]{mengel}Mengel, J.~G, Norris, J., 
\& Gross, P.~G. 1976, ApJ, 204, 488

\bibitem[Momany et al. (2002)]{momany}Momany, Y., Piotto, G., 
Recio-Blanco, A., Bedin, L.R., Cassisi, S., \& Bono, G. 2002, ApJ, 576, L65

\bibitem[Moni Bidin et al. (2006)]{bidin06}Moni Bidin, C., Moehler, S., Piotto, G., 
Recino-Blanco, A., Momany, Y., \& Mendez, R.~A. 2006, A\&A, 451, 499

\bibitem[Napiwotzki et al.(2004)]{napiwotzki}Napiwotzki, R., Karl, C.~A., 
Lisker, T., et a. 2004, Ap\&SS, 291, 321 

\bibitem[Pooley et al. (2002)]{pooley} Pooley, D., Lewin, W.~H.~G.,
Homer, L., et al. 2002, ApJ, 569, 405  

\bibitem[Schwarzenberg-Czerny (1996)]{alex}Schwarzenberg-Czerny, A. 1996,
ApJ, 460, L107 

\bibitem[Schwarzenberg-Czerny \& Beaulieu (2006)]{alex1}
Schwarzenberg-Czerny, A., \& Beaulieu, J.-Ph.  2006, MNRAS, 365, 165 

\bibitem[Stetson(1987)]{stetson}
 Stetson, P. B. 1987, PASP, 99, 191


\bibitem[Thompson et al. (1999)]{thompson}Thompson, I.~B, Kaluzny, J., Pych, W.,
\& Krzeminski W. 1999, AJ, 118, 462 


\end{thebibliography}
\end{document}